\documentclass[journal]{IEEEtran} 
\IEEEoverridecommandlockouts
\usepackage{setspace}
\usepackage{subcaption}
\usepackage{graphicx}
\usepackage{float}
\usepackage{amsmath}
\usepackage{amsmath,amssymb,amsfonts}
\usepackage{algorithm}
\usepackage{algorithmicx}
\usepackage{algpseudocode}
\usepackage{array}
\usepackage{amsthm}
\usepackage{amsmath}
\usepackage{amssymb}
 \usepackage{multirow}
\usepackage{mdwmath}
\usepackage{mdwtab}
\usepackage{eqparbox}
\usepackage{stfloats}
\usepackage{fixltx2e}
\usepackage{cases} 
\usepackage{caption}
\usepackage{graphicx}
\usepackage{float} 
\usepackage{booktabs} 
\usepackage{makecell} 

 \usepackage{flushend}

\usepackage{xcolor}
\usepackage{hyperref}
\hypersetup{
    colorlinks=true, 
    linkcolor=blue, 
    citecolor=blue, 
    urlcolor=blue 
}

\usepackage{url}
\usepackage{textcomp}
\usepackage{cite}
\def\BibTeX{{\rm B\kern-.05em{\sc i\kern-.025em b}\kern-.08em
    T\kern-.1667em\lower.7ex\hbox{E}\kern-.125emX}}
\allowdisplaybreaks[4]

\newtheorem{remark}{Remark}

\usepackage{wrapfig}

\makeatletter
\renewcommand{\maketag@@@}[1]{\hbox{\m@th\normalsize\normalfont#1}}%
\makeatother
\UseRawInputEncoding 

\makeatletter
\newcommand{\linebreakand}{%
  \end{@IEEEauthorhalign}
  \hfill\mbox{}\par
  \mbox{}\hfill\begin{@IEEEauthorhalign}
}
\makeatother

\begin{document}

\title
{GNN-Based Global CSI Reconstruction for Fronthaul-Limited Distributed MIMO Systems}
\author{ Haojin Li, Kaiqian Qu, Anbang Zhang, Chen Sun$^*$,~\IEEEmembership{Senior Member,~IEEE},  Wenqi Zhang,\\ Haijun Zhang,~\IEEEmembership{Fellow,~IEEE}
\thanks{
Haojin Li, and Haijun Zhang are with Beijing Engineering and Technology Research Center for Convergence Networks and Ubiquitous Services, University of Science and Technology Beijing, Beijing 100083, China (E-mail: zhanghaijun@ustb.edu.cn). $^{*}$ Corresponding author.

Haojin Li, Kaiqian Qu, Anbang Zhang, Chen Sun, Wenqi Zhang are with Wireless Network Research Department, Sony China Research Laboratory, Beijing, 100027, China (E-mail:\{haojin.li, chen.sun,  wenqi.zhang\}@sony.com). 
 }}

\maketitle


\begin{abstract}
	Global channel state information (CSI) acquisition is essential for cooperative precoding in distributed multiple-input multiple-output (DMIMO) systems, but uploading full instantaneous CSI from all distributed antennas creates heavy fronthaul overhead. This paper proposes a fronthaul-efficient acquisition framework based on graph neural network (GNN) reconstruction and task-driven antenna selection. Each transmission and reception point (TRP) uploads only selected antenna CSI, while the centralized unit (CU) reconstructs the full global CSI from partial observations. A universal mask-conditioned GNN is trained with random upload masks, used to evaluate antenna subsets under a fronthaul budget, and then fine-tuned for the selected deployment mask. Simulation results show improved CSI reconstruction accuracy with lower fronthaul and pilot overhead.
\end{abstract}


\section{Introduction}

Distributed multiple-input multiple-output (DMIMO), often studied in large-scale user-centric form as cell-free massive MIMO, is a promising architecture for beyond-fifth-generation (beyond-5G) and sixth-generation (6G) wireless networks. By replacing cell-centric transmission with geographically distributed transmission and reception points (TRPs) or access points (APs), DMIMO can provide macro-diversity, improve user fairness, and mitigate inter-cell interference~\cite{ngo2017cellfree}. These benefits become most pronounced when distributed antennas are coordinated through coherent joint processing, which in turn requires accurate global instantaneous channel state information (CSI) at the centralized unit (CU)~\cite{bjornson2020competitive}.

The acquisition and delivery of global instantaneous CSI is therefore a key scalability bottleneck. Existing cell-free massive MIMO studies have developed precoding and power-control methods~\cite{nayebi2017precoding}, centralized MMSE processing~\cite{bjornson2020competitive}, scalable user-centric architectures~\cite{9904601,bjornson2020scalable}, and local partial zero-forcing precoding~\cite{interdonato2020local}. These methods improve cooperative transmission under different processing assumptions, but they mainly focus on how to use CSI after it is obtained. They do not directly address the antenna-level acquisition problem considered here: how the CU can recover the global instantaneous CSI matrix when only a small subset of distributed antenna CSI vectors can be uploaded through limited fronthaul links.

Several overhead-reduction techniques are related but not fully suited to this setting. For FDD cell-free massive MIMO, angle-domain processing exploits long-term multipath structure to reduce CSI acquisition overhead~\cite{abdallah2020efficient}. Deep learning-based CSI feedback methods, such as CsiNet and CLNet, compress and reconstruct massive-MIMO CSI with neural autoencoder architectures~\cite{wen2018csinet,ji2021clnet}. However, such feedback methods usually require neural encoders at the feedback side and do not explicitly decide which physical antenna CSI vectors should be uploaded. Recent learning-based antenna selection has been jointly designed with precoding in cell-free MIMO networks~\cite{11084997}, but its objective is transmit-antenna activation for spectral-efficiency optimization rather than selecting uploaded CSI observations for global channel reconstruction.

To address this issue, this paper proposes a fronthaul-efficient global CSI acquisition framework. Each TRP only estimates and uploads the CSI of selected antenna dimensions, while all learning and reconstruction are performed at the CU. A universal mask-conditioned graph neural network (GNN) is first trained with random upload masks to recover global CSI from different partial observation patterns. The trained reconstructor is then used to perform task-driven antenna selection under a fronthaul budget, followed by selection-specific fine-tuning for the final deployment mask. Simulation results show that the proposed framework improves CSI reconstruction accuracy and downlink transmission performance while reducing fronthaul and pilot-based CSI acquisition overhead.

\section{System Model and Problem Formulation}
\label{sec:system_model}

\subsection{Fronthaul-Limited Distributed MIMO Architecture}

We consider a fronthaul-limited DMIMO system, where $M$ distributed TRPs are connected to a CU through fronthaul links. Each TRP is equipped with $N_t$ antennas, and the total number of distributed antennas is $N = MN_t$.
The CU is responsible for centralized cooperative signal processing, such as global CSI acquisition and joint transmission design. 

Let $\mathbf H_m \in \mathbb C^{N_r \times N_t}$ denote the instantaneous channel matrix from the $N_t$ antennas of the $m$-th TRP to a receiver equipped with $N_r$ antennas. The global CSI matrix is then given by
\begin{equation}
	\mathbf H =
	\left[
	\mathbf H_1,
	\mathbf H_2,
	\ldots,
	\mathbf H_M
	\right]
	\in \mathbb C^{N_r \times N},
	\label{eq:global_csi}
\end{equation}
where the $i$-th column of $\mathbf H$, denoted by $\mathbf h_i \in \mathbb C^{N_r}$, represents the channel vector associated with the $i$-th global antenna. The global antenna index $i$ can be mapped to a specific TRP and its corresponding local antenna index.

In conventional centralized DMIMO, each TRP uploads the instantaneous CSI, enabling the CU to obtain the full global CSI matrix $\mathbf H$. However, the resulting fronthaul overhead scales with the total number of distributed antennas and can become prohibitive as $M$ or $N_t$ increases. 

\begin{figure}[!t]
	\centering
	\includegraphics[width=0.8\linewidth]{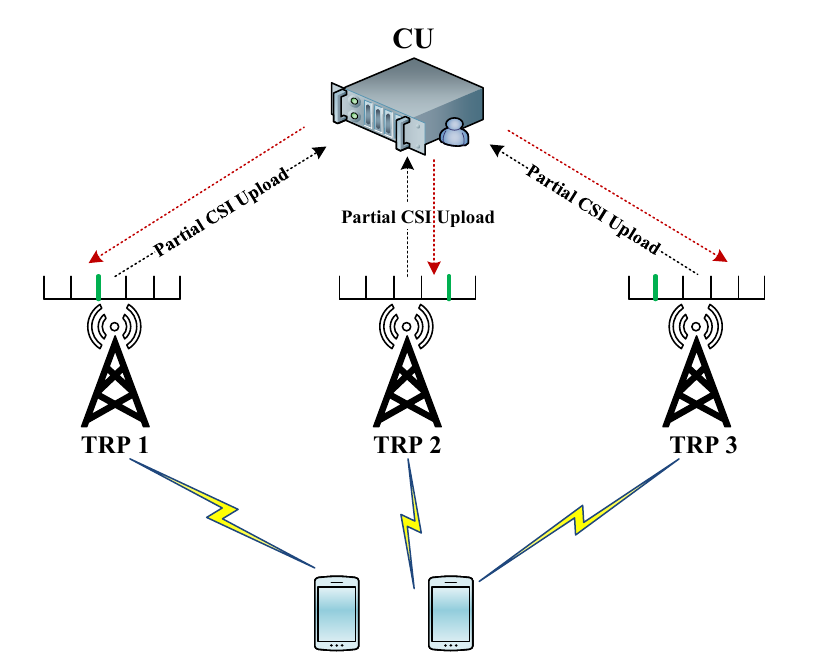}
	\caption{Illustration of the DMIMO system.}
	\label{fig:system_model}
\end{figure}


\subsection{Long-Term Spatial Channel Model}

The instantaneous global CSI is generated according to long-term spatial channel statistics. Specifically, we consider a correlated channel model
\begin{equation}
	\mathbf H
	=
	\mathbf R_r^{1/2}
	\mathbf W
	\mathbf R_t^{1/2},
	\label{eq:correlated_channel_model}
\end{equation}
where $\mathbf W \in \mathbb C^{N_r \times N}$ contains independent and identically distributed complex Gaussian entries with zero mean and unit variance, $\mathbf R_r \in \mathbb C^{N_r \times N_r}$ denotes the receive-side spatial correlation matrix, and $\mathbf R_t \in \mathbb C^{N \times N}$ denotes the transmit-side long-term covariance matrix across all distributed TRP antennas. For clarity and to highlight the transmit-side fronthaul compression problem, we set $\mathbf R_r=\mathbf I_{N_r}$.

The transmit-side covariance captures both large-scale fading and spatial correlation among distributed antennas. Its $(i,j)$-th entry can be expressed as
\begin{equation}
	[\mathbf R_t]_{i,j}
	=
	\sqrt{\beta_i\beta_j}\rho_{i,j},
	\label{eq:tx_covariance}
\end{equation}
where $\beta_i$ denotes the large-scale fading coefficient of the $i$-th global antenna, and $\rho_{i,j}$ represents the normalized long-term spatial correlation between antennas $i$ and $j$. A commonly used exponential spatial correlation model is adopted to
describe the distance-dependent long-term correlation among distributed
antennas~\cite{choi2014bounds}. Specifically,
we set
\begin{equation}
	\rho_{i,j}
	=
	\exp\left(-\frac{d_{i,j}}{d_c}\right),
	\label{eq:distance_correlation}
\end{equation}
where $d_{i,j}$ is the physical distance between the two antennas and $d_c$ is the correlation distance parameter.

The long-term channel statistics are assumed to vary much more slowly than the instantaneous CSI. Therefore, they can be estimated or updated over a long time scale and made available at the CU. In contrast, the instantaneous CSI needs to be uploaded frequently and hence dominates the fronthaul signaling overhead. 

\subsection{Partial CSI Upload and Global CSI Reconstruction} Fortunately, the long-term spatial correlation among TRP antennas makes it possible to infer the unuploaded antenna CSI from a small subset of observed antenna CSI. Let
$
	\mathcal S \subseteq \{1,2,\ldots,N\}$
denote the selected global antenna index set for CSI upload. The total number of uploaded antenna CSI vectors is limited by
$
	|\mathcal S| = K,\ K \ll N .
$
The corresponding partial CSI observed by the CU is
\begin{equation}
	\mathbf H_{\mathcal S}
	=
	\mathbf H(:,\mathcal S).
	\label{eq:partial_csi}
\end{equation}
The uploaded CSI corresponds to the observed columns of $\mathbf H$, while the remaining columns are unavailable at the CU and need to be reconstructed.
\begin{remark}[Acquisition of the uploaded CSI]
	The uploaded partial CSI $\mathbf H_{\mathcal S}$ is obtained through conventional pilot-aided least-squares (LS) channel estimation at the selected antenna dimensions. Since only $K$ antenna CSI vectors are estimated and uploaded, where $K\ll N$, the LS estimation involves only a low-dimensional channel matrix and therefore has very low computational complexity. 
\end{remark}

The objective of global CSI reconstruction is to recover the full CSI matrix from the partial upload
\begin{equation}
	\hat{\mathbf H}
	=
	\mathcal R
	\left(
	\mathbf H_{\mathcal S},\mathcal S;\boldsymbol{\Theta}
	\right),
	\label{eq:reconstruction_mapping}
\end{equation}
where $\mathcal R(\cdot)$ denotes a reconstruction function deployed at the CU and $\boldsymbol{\Theta}$ represents its learnable parameters.

This problem is challenging for several reasons. The key difficulty is that partial CSI upload introduces a coupled reconstruction and selection problem. On the one hand, the CU has to infer the missing antenna CSI from only a small number of uploaded CSI vectors. On the other hand, the reconstruction quality depends strongly on which antenna CSI vectors are uploaded. Therefore, the problem is not only how to reconstruct the global CSI from partial observations, but also how to determine an informative subset of antenna CSI for upload.

\section{Proposed GNN-Based CSI Reconstruction Framework}
\label{sec:proposed_framework}


The proposed framework consists of three stages: universal GNN reconstructor training, task-driven antenna selection, and selection-specific fine-tuning. 
In the first stage, the CU trains a universal mask-conditioned GNN reconstructor using randomly generated antenna upload masks subject to the fronthaul budget. For each mask, only the CSI of selected antennas is treated as observed, while the the CSI from the remaining antennas is masked out.

In the second stage, the trained universal reconstructor is used as an evaluator for antenna upload selection. Since different antenna subsets may lead to different reconstruction quality, the CU searches for an informative subset of antennas under the upload budget. The selection criterion can be based on validation reconstruction error metric. 


In the third stage, once the final antenna upload set is obtained, the universal reconstructor is fine-tuned using this fixed selected mask. This step adapt s the reconstructor from a general mask-conditioned model to a selection-specific model, thereby improving reconstruction quality for the final deployment pattern.


\subsection{GNN-Based Global CSI Reconstruction}
\label{subsec:gnn_reconstruction}
\begin{figure}[!t]
	\centering
	\includegraphics[width=1\linewidth]{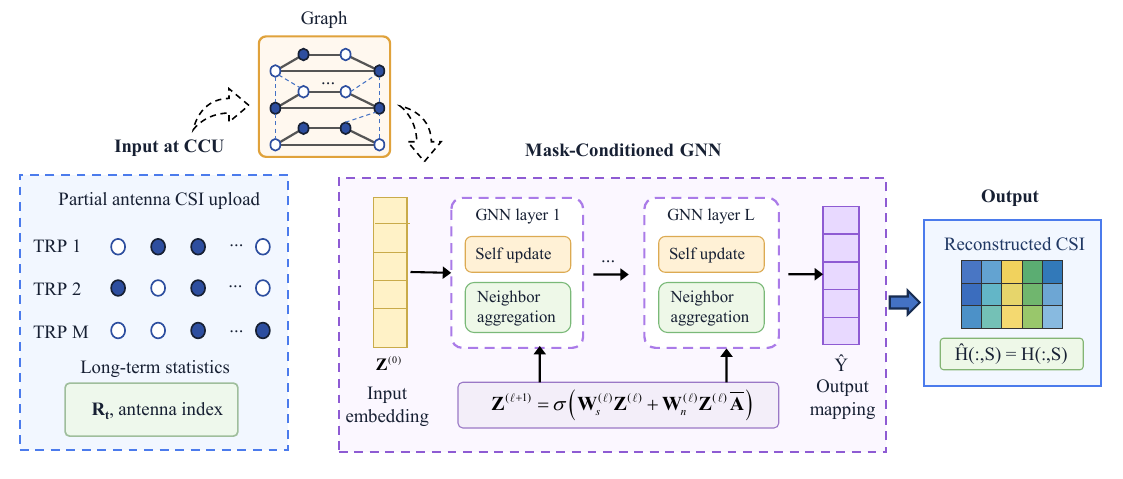}
	\caption{GNN-based CSI reconstruction.}
	\label{fig:gnn_graph}
\end{figure}
The partial CSI upload problem can be naturally interpreted as a graph signal completion problem. Specifically, the instantaneous CSI associated with each distributed antenna is regarded as a node signal, while the long-term statistical dependence among antennas provides the graph structure. Under this view, the CU reconstructs the missing antenna CSI by propagating information from uploaded antenna nodes to unuploaded antenna nodes over an antenna-level graph~\cite{10713240}.

\subsubsection{Antenna-Level Graph Representation}
\label{subsubsec:graph_representation}

We construct an antenna-level graph
\begin{equation}
	\mathcal G = (\mathcal V,\mathcal E,\mathbf A),
	\label{eq:antenna_graph}
\end{equation}
where $\mathcal V={1,2,\ldots,N}$ denotes the set of antenna nodes, $\mathcal E$ denotes the set of graph edges, and $\mathbf A\in\mathbb R^{N\times N}$ is the adjacency matrix. This antenna-level modeling is important because the fronthaul constraint is imposed on the number of uploaded antenna CSI vectors.

For the $i$-th antenna node, let $\mathbf h_i=\mathbf H(:,i)\in\mathbb C^{N_r}$ denote its instantaneous CSI vector and $m_i\in{0,1}$ denote its upload indicator. If antenna $i$ is uploaded, i.e.,
$m_i=1,$ the CU observes $\mathbf h_i$; otherwise, the instantaneous CSI of this node is unavailable. The observed CSI vector can be written as
\begin{equation}
	\widetilde{\mathbf h}_i = m_i \mathbf h_i .
	\label{eq:observed_node_csi}
\end{equation}

The complex CSI is represented by its real and imaginary parts. The node feature of antenna $i$ is constructed as
\begin{equation}
	\mathbf x_i =
	\left[
	\mathrm{Re}(\widetilde{\mathbf h}_i)^T,
	\mathrm{Im}(\widetilde{\mathbf h}_i)^T,
	m_i,
	q_i
	\right]^T,
	\label{eq:node_feature_method}
\end{equation}
where $ q_i$ denotes the local antenna index. For unuploaded antennas, the instantaneous CSI entries in $\mathbf x_i$ are set to zero, while the mask entry $m_i=0$ explicitly informs the GNN that the corresponding CSI is missing.

Stacking all node features gives the graph input matrix
\begin{equation}
	\mathbf X(\mathcal S)
	=
	\left[
	\mathbf x_1,
	\mathbf x_2,
	\ldots,
	\mathbf x_N
	\right]
	\in \mathbb R^{F\times N},
	\label{eq:graph_input_matrix}
\end{equation}
where $F$ is the node feature dimension. The reconstruction task is to infer the missing node signals ${\mathbf h_i:i\notin\mathcal S}$ from the observed node signals ${\mathbf h_i:i\in\mathcal S}$, the upload mask, and the antenna graph.

\subsubsection{Long-Term-Correlation-Guided Graph Construction}
\label{subsubsec:graph_construction}

The graph adjacency matrix should reflect the long-term statistical dependence among antenna nodes. A direct choice would be to use the transmit-side covariance matrix $\mathbf R_t$. However, the raw covariance coefficient $[\mathbf R_t]_{i,j}$ contains both spatial correlation and large-scale fading effects. Directly using $|[\mathbf R_t]_{i,j}|$ as an edge weight may bias the graph toward high-power antennas rather than antennas that are truly more correlated.

To avoid this issue, we first normalize the covariance into a correlation coefficient:
\begin{equation}
	c_{i,j}
	=
	\frac{|[\mathbf R_t]_{i,j}|}
	{\sqrt{[\mathbf R_t]_{i,i}[\mathbf R_t]_{j,j}}}.
	\label{eq:normalized_correlation_method}
\end{equation}

In distributed deployments, antennas within the same TRP usually exhibit stronger useful correlation than antennas belonging to different TRPs. Therefore, instead of constructing a fully connected global graph, we adopt a TRP-wise local graph to suppress irrelevant cross-TRP message propagation. Let $\mathcal A_m$ denote the set of global antenna indices belonging to the $m$-th TRP. The adjacency matrix is constructed as
\begin{equation}
	[\mathbf A]_{i,j}
	=
	\begin{cases}
		c_{i,j}, & i,j\in\mathcal A_m,\ \exists m,\ i\neq j,\\
		0, & \text{otherwise}.
	\end{cases}
	\label{eq:trp_wise_adjacency}
\end{equation}

Self-loops are then added to preserve each node's own information, and symmetric normalization is applied:
\begin{equation}
	\overline{\mathbf A}
	=
	\mathbf D^{-\frac{1}{2}}
	\left(
	\mathbf A+\mathbf I
	\right)
	\mathbf D^{-\frac{1}{2}},
	\label{eq:normalized_adjacency}
\end{equation}
where $\mathbf D$ is the degree matrix of $\mathbf A+\mathbf I$. The normalized adjacency matrix $\overline{\mathbf A}$ is used for graph message passing.

\subsubsection{Mask-Conditioned GNN Architecture}
\label{subsubsec:gnn_architecture}

The proposed reconstructor is a mask-conditioned GNN deployed at the CU. Its input is the graph feature matrix $\mathbf X(\mathcal S)$ and the normalized adjacency matrix $\overline{\mathbf A}$, and its output is the reconstructed global CSI matrix $\hat{\mathbf H}$.

First, the input node features are mapped into a hidden representation:
\begin{equation}
	\mathbf Z^{(0)}
	=
	\phi_{\mathrm{in}}
	\left(
	\mathbf X(\mathcal S)
	\right),
	\label{eq:input_embedding}
\end{equation}
where $\phi_{\mathrm{in}}(\cdot)$ denotes a learnable feature embedding function. Then, $L$ graph message-passing layers are used to propagate information over the antenna graph. The $\ell$-th message-passing layer is given by
\begin{equation}
	\mathbf Z^{(\ell+1)}
	=
	\sigma
	\left(
	\mathbf W_s^{(\ell)}\mathbf Z^{(\ell)}
	+
	\mathbf W_n^{(\ell)}
	\mathbf Z^{(\ell)}
	\overline{\mathbf A}
	\right),
	\
	\ell=0,1,\ldots,L-1,
	\label{eq:gnn_message_passing}
\end{equation}
where $\mathbf Z^{(\ell)}$ is the node representation at the $\ell$-th layer, $\mathbf W_s^{(\ell)}$ and $\mathbf W_n^{(\ell)}$ are learnable weights for self-node transformation and neighbor aggregation, respectively, and $\sigma(\cdot)$ is a nonlinear activation function. The term $\mathbf Z^{(\ell)}\overline{\mathbf A}$ aggregates information from statistically related antenna nodes.

After message passing, an output mapping function predicts the real and imaginary parts of the global CSI:
\begin{equation}
	\hat{\mathbf Y}
	=
	\phi_{\mathrm{out}}
	\left(
	\mathbf Z^{(L)}
	\right),
	\label{eq:gnn_output_mapping}
\end{equation}
where $\hat{\mathbf Y}\in\mathbb R^{2N_r\times N}$ contains the predicted real and imaginary CSI components of all antenna nodes. The reconstructed complex CSI associated with antenna $i$ is obtained as
\begin{equation}
	\hat{\mathbf h}_i
	=
	\hat{\mathbf y}_{i,\mathrm R}
	+
	j\hat{\mathbf y}_{i,\mathrm I},
	\quad i=1,2,\ldots,N,
	\label{eq:complex_output}
\end{equation}
where $\hat{\mathbf y}_{i,\mathrm R}$ and $\hat{\mathbf y}_{i,\mathrm I}$ are the predicted real and imaginary components associated with antenna $i$.

To ensure that the uploaded CSI is exactly preserved, data consistency is enforced after the GNN output:
\begin{equation}
	\hat{\mathbf H}(:,i)
	=
	\begin{cases}
		\mathbf H(:,i), & i\in\mathcal S,\\
		\hat{\mathbf H}_{\mathrm{GNN}}(:,i), & i\notin\mathcal S,
	\end{cases}
	\label{eq:data_consistency_method}
\end{equation}
where $\hat{\mathbf H}_{\mathrm{GNN}}$ denotes the raw GNN output. Therefore, the GNN only needs to infer the unuploaded antenna CSI, while the actually uploaded CSI remains unchanged.

\subsubsection{Universal Reconstructor Training}
\label{subsubsec:universal_training}

We first train a universal mask-conditioned reconstructor using randomly generated upload masks. During training, for each mini-batch, an antenna subset $\mathcal S$ is sampled under the fronthaul budget
$
	|\mathcal S|=K.$
To avoid losing the instantaneous CSI of an entire TRP, the random mask can be generated in a balanced manner such that each TRP contributes at least one uploaded antenna when $K\ge M$.

For each sampled mask, the input feature matrix $\mathbf X(\mathcal S)$ is constructed by zero-filling the unuploaded CSI entries and appending the corresponding mask indicators. The GNN then reconstructs the full global CSI. The universal reconstruction loss is computed over the missing antenna columns:
\begin{equation}
	\mathcal L_{\mathrm{rec}}(\theta)
	=
	\mathbb E_{\mathbf H,\mathcal S}
	\left[
	\frac{
		\left\|
		\hat{\mathbf H}(:,\mathcal S^c)
		-
		\mathbf H(:,\mathcal S^c)
		\right\|_F^2
	}{
		\left\|
		\mathbf H(:,\mathcal S^c)
		\right\|_F^2
		+\epsilon
	}
	\right],
	\label{eq:universal_training_loss}
\end{equation}
where $\epsilon$ is a small positive constant for numerical stability. The trained universal reconstructor will then serve as a stable evaluator for task-driven antenna selection.

\subsection{Task-Driven Antenna Selection and Selection-Specific Fine-Tuning}
\label{subsec:selection_finetuning}


Let $\theta_0$ denote the parameters of the trained universal reconstructor. For a candidate antenna set $\mathcal S$, the validation reconstruction loss is written as
\begin{equation}
	\mathcal L_{\mathrm{val}}
	\left(
	\mathcal S;\theta_0
	\right)
	=
	\frac{1}{|\mathcal D_{\mathrm{val}}|}
	\sum_{\mathbf H\in\mathcal D_{\mathrm{val}}}
	\frac{
		\left\|
		\hat{\mathbf H}_{\theta_0}(:,\mathcal S^c)
		-
		\mathbf H(:,\mathcal S^c)
		\right\|_F^2
	}{
		\left\|
		\mathbf H(:,\mathcal S^c)
		\right\|_F^2
		+\epsilon
	},
	\label{eq:validation_loss_selection}
\end{equation}
where $\mathcal D_{\mathrm{val}}$ denotes the validation set and $\hat{\mathbf H}_{\theta_0}$ denotes the reconstruction obtained by the universal reconstructor under the candidate upload set $\mathcal S$.

The task-driven antenna selection problem is formulated as
\begin{equation}
	\mathcal S^{\star}
	=
	\arg\min_{\mathcal S}
	\mathcal L_{\mathrm{val}}
	\left(
	\mathcal S;\theta_0
	\right),
	\label{eq:task_driven_selection}
\end{equation}
subject to $
	|\mathcal S|=K.$
To avoid losing the instantaneous CSI of an entire TRP, we further impose a TRP coverage constraint:
\begin{equation}
	|\mathcal S\cap\mathcal A_m|\ge 1,
	\quad m=1,2,\ldots,M.
	\label{eq:trp_coverage_constraint}
\end{equation}

The exact solution to \eqref{eq:task_driven_selection} requires combinatorial search over all possible antenna subsets and is generally impractical when $N$ is large. Therefore, an efficient TRP-seeded greedy strategy is adopted. First, one representative antenna is selected from each TRP to satisfy the TRP coverage constraint. Then, the remaining upload budget is allocated by iteratively adding the antenna that yields the largest reduction in the validation loss. Specifically, if $\mathcal S^{(t)}$ denotes the selected set at the $t$-th iteration, the next antenna is chosen as
\begin{equation}
	i^{\star}
	=
	\arg\min_{i\notin\mathcal S^{(t)}}
	\mathcal L_{\mathrm{val}}
	\left(
	\mathcal S^{(t)}\cup{i};\theta_0
	\right).
	\label{eq:greedy_selection_step}
\end{equation}
The selected set is updated as
\begin{equation}
	\mathcal S^{(t+1)}
	=
	\mathcal S^{(t)}\cup{i^{\star}}.
	\label{eq:greedy_selection_update}
\end{equation}
This process continues until $|\mathcal S^{(t)}|=K$, and the resulting set is used as the final upload antenna set $\mathcal S^{\star}$.

After obtaining $\mathcal S^{\star}$, the universal reconstructor is fine-tuned with this fixed antenna mask. The selection-specific reconstructor is obtained as
\begin{equation}
	\theta_{\mathcal S^{\star}}
	=
	\mathrm{FineTune}
	\left(
	\theta_0,\mathcal S^{\star}
	\right).
	\label{eq:selection_specific_finetuning}
\end{equation}
The final reconstructed global CSI is then given by
\begin{equation}
	\hat{\mathbf H}
	=
	f_{\theta_{\mathcal S^{\star}}}
	\left(
	\mathbf X(\mathcal S^{\star}),
	\overline{\mathbf A}
	\right).
	\label{eq:final_reconstruction}
\end{equation}

\section{Experimental Results and Analysis}
\label{sec:experiments}

This section evaluates the proposed framework by simulation. We consider $M=4$ TRPs with $N_t=16$ antennas each, $N_r=2$ receive antennas, and $K=4$ uploaded antenna CSI vectors unless otherwise specified. In total, $5000$ channel samples are divided into training, validation, and test sets with a ratio of $0.8:0.1:0.1$. The GNN has hidden dimension $128$ and $L=2$ message-passing layers; it is trained for $80$ epochs and fine-tuned for $30$ epochs. The normalized mean-squared error (NMSE) and cumulative distribution function (CDF) are adopted as the performance metrics. Monte Carlo evaluation uses $1000$ independent samples.
\footnote{The hidden dimension of GNN is $128$, and the network contains $L=2$ graph message-passing layers followed by a node-wise output head.  The universal GNN is trained for $80$ epochs with mini-batch size $128$ and evaluation batch size $256$. Adam is adopted with initial learning rate $10^{-4}$, decay factors $0.9$ and $0.999$, and gradient clipping threshold $0.5$. The learning rate is halved after $60$ epochs. For task-driven selection, $128$ validation samples are used. Selection-specific fine-tuning is performed for $30$ epochs with learning rate $2\times10^{-5}$. The convolutional neural network (CNN) baseline uses the same data split, selection, and fine-tuning pipeline, but replaces the GNN with a plain CNN.}

\begin{figure}[!t]
	\centering
	\begin{minipage}{0.48\linewidth}
		\centering
		\includegraphics[width=\linewidth]{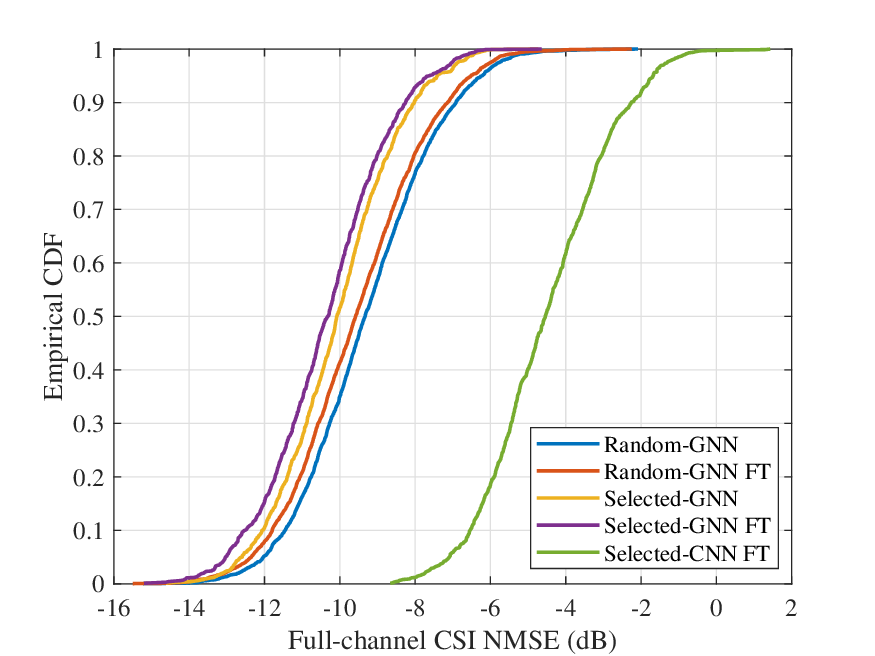}
		\captionof{figure}{CDF of CSI reconstruction NMSE.}
		\label{fig:nmse_cdf}
	\end{minipage}
	\hfill
	\begin{minipage}{0.48\linewidth}
		\centering
		\includegraphics[width=\linewidth]{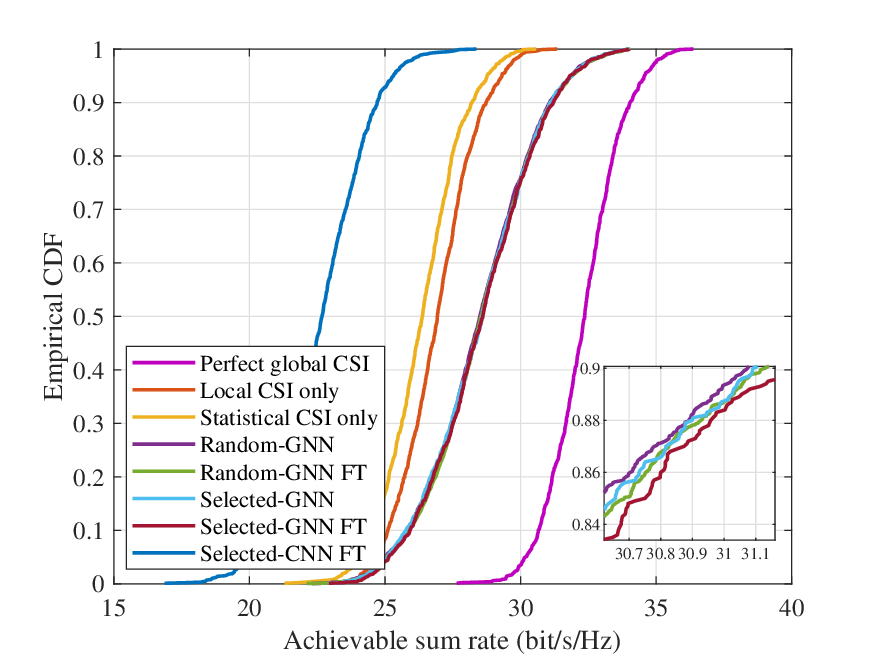}
		\captionof{figure}{CDF of achievable sum rate.}
		\label{fig:sumrate_cdf}
	\end{minipage}
\end{figure}

The compared schemes are summarized as follows. \textbf{Random-GNN} uses randomly uploaded antenna CSI with the universal GNN reconstructor, and \textbf{Random-GNN-FT} further fine-tunes the reconstructor under the fixed random mask. \textbf{Selected-GNN} uses the task-driven antenna subset selected by the universal reconstructor, while \textbf{Selected-GNN-FT} further performs selection-specific fine-tuning and denotes the complete proposed method. \textbf{Selected-CNN-FT} follows the same pipeline but replaces the GNN with a plain CNN. For communication-rate comparison, all schemes design the transmit precoder from the reconstructed global CSI using singular value decomposition (SVD) with water-filling power allocation, and the achievable sum rate is evaluated over the true instantaneous channel.

Fig.~\ref{fig:nmse_cdf} shows the empirical CDF of the CSI reconstruction NMSE for $K=4$ at signal-to-noise ratio (SNR) $=20$ dB. The task-driven selection schemes achieve better NMSE distributions than random upload, indicating that the selected antenna subset is critical for global CSI recovery. The universal reconstructor serves as an evaluator and selects antennas that are more informative for recovering the complete channel. Selected-GNN-FT further improves over Selected-GNN, because the fixed-mask fine-tuning stage adapts the model to the final deployment pattern. Random-GNN-FT also improves over Random-GNN, but remains inferior to task-driven selection. The CNN baseline performs worse than the GNN schemes, confirming the benefit of graph message passing for exploiting antenna-level spatial correlation.

Fig.~\ref{fig:sumrate_cdf} presents the achievable rate CDF for $K=4$ at SNR $=20$ dB. Perfect global CSI provides the upper bound. Statistical CSI only performs poorly because it ignores instantaneous CSI, while local per-TRP CSI also suffers since independent TRP precoding cannot fully exploit distributed spatial degrees of freedom. Selected-GNN and Selected-GNN-FT outperform random upload, consistent with the NMSE results. Among all fronthaul-limited schemes, Selected-GNN-FT achieves the best sum-rate distribution and is closest to the perfect-CSI upper bound, showing that the proposed three-stage design improves both CSI recovery and transmission performance.

\begin{figure}[!t]
	\centering
	\begin{minipage}{0.48\linewidth}
		\centering
		\includegraphics[width=\linewidth]{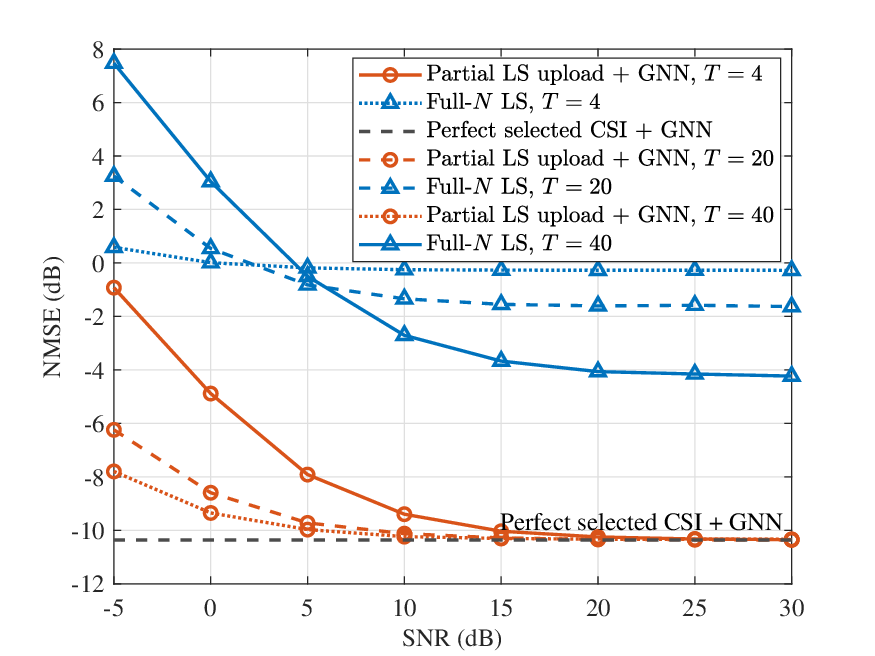}
		\captionof{figure}{LS channel-estimation NMSE versus SNR.}
		\label{fig:ls_snr_nmse}
	\end{minipage}
	\hfill
	\begin{minipage}{0.48\linewidth}
		\centering
		\includegraphics[width=\linewidth]{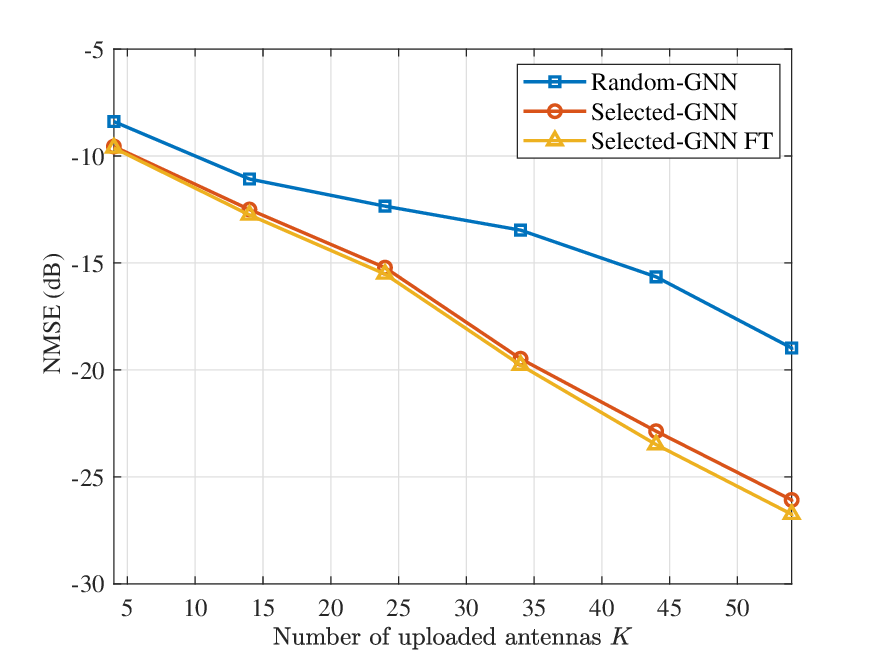}
		\captionof{figure}{Reconstruction NMSE versus upload budget $K$.}
		\label{fig:nmse_vs_k}
	\end{minipage}
\end{figure}

Fig.~\ref{fig:ls_snr_nmse} evaluates the effect of LS estimation error. For the LS estimation, the pilot length is denoted by $T$. The full-$N$ LS baseline estimates all $N$ antenna CSI vectors under the same pilot length, whereas partial LS upload estimates only the selected antennas. When pilot resources are limited, partial LS upload gives more reliable observations because it solves a lower-dimensional estimation problem. As the pilot SNR increases, the NMSE of partial LS upload with GNN reconstruction decreases and approaches the perfect-selected-CSI reference. The gain is more evident for shorter pilots, showing that the proposed method reduces both fronthaul load and CSI acquisition burden.

Fig.~\ref{fig:nmse_vs_k} shows NMSE versus the upload budget $K$. NMSE decreases as $K$ increases because more instantaneous CSI observations are available at the CU. Under the same $K$, Selected-GNN consistently outperforms Random-GNN, demonstrating that task-driven selection provides more informative observations under the same fronthaul overhead.

\section{Conclusion}

This paper proposed a fronthaul-efficient global CSI acquisition framework for DMIMO systems. By combining universal mask-conditioned GNN reconstruction, task-driven antenna selection, and selection-specific fine-tuning, the proposed method reconstructs global CSI from limited uploaded antenna CSI without deploying neural encoders at TRPs. Simulation results verified its advantages in CSI reconstruction accuracy, pilot-efficiency, and adaptability to different fronthaul budgets.




\bibliographystyle{IEEEtran} 
\bibliography{IEEEabrv,bib}
\end{document}